\documentstyle[12pt]{article}
\begin{document}
\begin{flushright}
BHU-PHYS-CAS Preprint\\
arXiv: 1005.5067 [hep-th]
\end{flushright}
\vskip 2cm
\begin{center}
{\Large {\sf Nilpotent (anti-)BRST symmetry transformations for dynamical 
non-Abelian 2-form gauge theory: superfield formalism}}

\vskip 3cm

{\large {\sf R. P. Malik}}\\
{\it Physics Department, Centre of Advanced Studies,}\\
{\it Banaras Hindu University, Varanasi - 221 005, (U.P.), India}\\

\vskip 0.1cm

{\bf and}\\

\vskip 0.1cm

{\it DST Centre for Interdisciplinary Mathematical Sciences,}\\
{\it Faculty of Science, Banaras Hindu University, Varanasi - 221 005, India}\\
{\sf {e-mails: rudra.prakash@hotmail.com ; malik@bhu.ac.in}}

\end{center}

\vskip 2cm

\noindent
{\bf Abstract:} We derive the off-shell nilpotent and absolutely anticommuting
Becchi-Rouet-Stora-Tyutin (BRST) and anti-BRST symmetry transformations for the
dynamical four (3 + 1)-dimensional (4D)
non-Abelian 2-form gauge theory within the framework of geometrical superfield formalism.
We obtain the (anti-) BRST invariant coupled Lagrangian densities that respect the
above nilpotent symmetry transformations. We discuss, furthermore, this (anti-) BRST invariance 
in the language of the superfield formalism. One of the novel features of our present investigation
is the observation that, in addition to the horizontality condition, we 
are theoretically compelled to invoke some other
physically relevant restrictions in order to deduce the precise (anti-) BRST symmetry transformations
for {\it all} the fields of a topologically massive 4D non-Abelian gauge theory.\\

$~~~~~~~~~~~~~~~~~~~~~~~$11.15.-q$~~~~~\;$  Gauge field theories\\
$~~~~~~~~~~~~~~~~~~~~~~~$03.70.+k $~~~$ Theory of quantized fields\\



\newpage

\noindent
\section {Introduction}


The central theme of our present investigation is to exploit the potential and power
of the superfield formalism (see, e.g. [1]), that has been successfully applied in the
context of (non-)Abelian 1-form, Abelian 2-form and 3-form gauge theories (see, e.g. [2-4]),
in the description of the four (3 + 1)-dimensional (4D) topologically
massive non-Abelian gauge theory where there is an explicit coupling between the non-Abelian 2-form
($B^{(2)} = \frac{1}{2!} (dx^\mu \wedge dx^\nu) B_{\mu\nu}$) gauge field $B_{\mu\nu} = B_{\mu\nu}\cdot T$ 
and the non-Abelian 1-form ($A^{(1)} = dx^\mu A_\mu $)
gauge field $A_\mu = A_\mu \cdot T$ through the famous topological ($B^{(2)} \wedge F^{(2)}$) term where the
2-form $F^{(2)} = \frac{1}{2!} (dx^\mu \wedge dx^\nu) F_{\mu\nu}$ defines the curvature tensor
$F_{\mu\nu} = \partial_\mu A_\nu - \partial_\nu A_\mu + i [A_\mu, A_\nu]$ corresponding to the
1-form gauge potential $A_\mu$. Here all the gauge fields are defined
in the adjoint representation of the semi-simple non-Abelian gauge group $SU(N)$.

Since the Higgs particles (that are responsible for generating masses
for the gauge particles and fermions in the domain of 
standard model of high energy physics)
have not yet been observed experimentally, it has become an issue of paramount importance
to construct gauge-invariant theories that could provide masses to the gauge particles
and fermions without taking any recourse to the Higgs mechanism. In this context, the study of
4D topologically massive gauge theories of Abelian and non-Abelian types has become
quite popular because the latter do provide a theoretical basis for generating masses
for the gauge bosons without exploiting any inputs from the Higgs mechanism (see, e.g. [5]).

Recently, we have studied the 4D topologically massive Abelian gauge theory within the framework 
of BRST formalism [6]. Its straightforward generalization to the non-Abelian
topologically massive theory is non-trivial because of some very strong no-go theorems [7]. There
are, at least, a couple of models [8,9], however, that circumvent the severe strictures 
laid down by the above no-go theorems. In  our
present endeavor, we shall focus on the dynamical non-Abelian 2-form gauge theory [9] and
study its BRST and anti-BRST structures by exploiting its {\it usual} 
``scalar'' gauge symmetry transformations within the 
framework of the geometrical superfield formalism proposed in [1-3].

One of the highlights of our findings is that the
gauge-invariant restrictions are invoked, in addition to the horizontality condition, for
the exact derivation of {\it all} the off-shell nilpotent and
absolutely anticommuting (anti-) BRST transformations. This observation, 
to the best of our knowledge, is a new result within the framework of the application of 
superfield formalism to a gauge theory (without any interaction with matter fields).

Let us begin with the Lagrangian density for the four (3 + 1)-dimensional (4D)
topologically massive non-Abelian gauge theory\footnote{We adopt here the conventions and notations 
such that the background 4D Minkowski spacetime manifold has the flat metric with signatures
(+1, -1, -1, -1) and the group generators $T^a$ of the $SU(N)$ group obey the Lie algebra
$[T^a, T^b ] = i f^{abc} T^c$ with structure constants $f^{abc}$ (that are chosen to be
totally antisymmetric in indices $a, b,$ and $c$ where $a, b, c....= 1, 2, .....N^2 -1$). In the algebraic space, we also have: $(V \cdot W) = V^a W^a$ and $(V \times W)^a = f^{abc} V^b W^c$ for the sake of brevity.
The 4D Levi-Civita tensor $\varepsilon_{\mu\nu\eta\kappa}$ 
(with $\mu, \nu, \eta...= 0, 1, 2, 3$) satisfies
$\varepsilon_{\mu\nu\eta\kappa} \varepsilon^{\mu\nu\eta\kappa} = - 4!, \varepsilon_{\mu\nu\eta\kappa} \varepsilon^{\mu\nu\eta\sigma} = - 3! \delta^\sigma_\kappa$, etc., and $\varepsilon_{0123} = +1$.} (see, e.g. [9] 
for details)
\begin{eqnarray}
{\cal L}_0 = - \frac{1}{4}\; F^{\mu\nu} \cdot F_{\mu\nu} + \frac{1}{12} \;H^{\mu\nu\eta} \cdot H_{\mu\nu\eta}
+ \frac{m}{4} \;\varepsilon_{\mu\nu\eta\kappa} B^{\mu\nu} \cdot F^{\eta\kappa},
\end{eqnarray}
where the 2-form $F^{(2)} = d A^{(1)} + i \;A^{(1)} \wedge A^{(1)} \equiv 
\frac{1}{2!} (dx^\mu \wedge dx^\nu) F_{\mu\nu} \cdot T$ defines the curvature tensor
$F_{\mu\nu}$ for the gauge potential $A_\mu$, 3-form $H^{(3)} = \frac{1}{3!}\;
(dx^\mu \wedge dx^\nu \wedge dx^\eta)\; H_{\mu\nu\eta} \cdot T$ defines the compensated curvature tensor
in terms of the dynamical 2-form gauge potential
$B_{\mu\nu}$ and 1-form ($K^{(1)} =  dx^\mu \;K_\mu \cdot T$) auxiliary field $K_\mu$ 
as
\begin{eqnarray}
H^a_{\mu\nu\eta} &=& (\partial_\mu B^a_{\nu\eta} + \partial_\nu B^a_{\eta\mu} + \partial_\eta B^a_{\mu\nu})
- \bigl [(A_\mu \times B_{\nu\eta})^a + (A_\nu \times B_{\eta\mu})^a + (A_\eta \times B_{\mu\nu})^a \bigr ] \nonumber\\
&-& \bigl [(K_\mu \times F_{\nu\eta})^a + (K_\nu \times F_{\eta\mu})^a + (K_\eta \times F_{\mu\nu})^a \bigr ],
\end{eqnarray}
and the last term in the above Lagrangian density (1) corresponds to the topological mass term where the curvature
tensor $F_{\mu\nu}$ corresponding to the 
non-Abelian 1-form gauge field and the dynamical 2-form gauge field $B_{\mu\nu}$
are coupled together through $B^{(2)}\wedge F^{(2)}$.

The above Lagrangian density respects the {\it usual} infinitesimal ``scalar''
gauge transformations $\delta_g $ corresponding to the non-Abelian 1-form gauge theory as (see, e.g. [9])
\begin{eqnarray}
&\delta_g A_\mu = D_\mu \Omega \equiv \partial_\mu \Omega - (A_\mu \times \Omega), \qquad
\delta_g F_{\mu\nu} = - (F_{\mu\nu} \times \Omega), \qquad  
\delta_g B_{\mu\nu} = - (B_{\mu\nu} \times \Omega),& \nonumber\\
& \delta_g H_{\mu\nu\eta} = - (H_{\mu\nu\eta} \times \Omega), \qquad \delta_g K_\mu = - (K_\mu \times \Omega),
\quad \delta_g {\cal L}_0 = 0,
\end{eqnarray}
where $\Omega = \Omega \cdot T$ is the infinitesimal
$SU(N)$-valued ``scalar'' gauge parameter. In addition, there exists an
independent ``vector'' gauge symmetry transformation in the theory 
[9]. We shall exploit, however, the usual 
``scalar'' gauge symmetry
transformations (3) (and corresponding properties of the gauge-invariance) for our present discussion
of the 4D topologically massive gauge theory
 within the framework of superfield approach [1-3].


Our present paper is organized as follows. In Sec. 2, we recapitulate the bare essentials 
of the superfield approach [1-3] to derive the (anti-) BRST symmetry transformations
and Curci-Ferrari (CF) restriction [10]
for the non-Abelian 1-form gauge theory where the horizontality condition (HC) plays a decisive role. 
Our Sec. 3 is devoted to the derivation of (anti-) BRST symmetry transformations
for the non-Abelian 2-form gauge field and 1-form auxiliary field by exploiting a couple of physically
relevant restrictions that are distinctly different from the HC. We discuss, in Sec. 4,
the (anti-) BRST invariance of the topologically massive non-Abelian gauge theory that is 
described by the coupled Lagrangian densities. 
Finally, in Sec. 5, we summarize our results and make some concluding remarks.

\section{Nilpotent symmetry transformations for the non-Abelian 1-form gauge theory: superfield formalism}

In the superfield approach to BRST formalism [1-3], one generalizes the 4D basic non-Abelian gauge
field ($A_\mu = A_\mu \cdot T$) and fermionic (anti-) ghost fields ($\bar C = \bar C \cdot T, C = C \cdot T$)
to the superfields defined on the (4, 2)-dimensional supermanifold. These superfields are expanded along
the Grassmannian directions of the supermanifold as (see, e.g. [1,2])
\begin{eqnarray}
{\cal B}_\mu (x, \theta, \bar \theta) &=& A_\mu (x) + \theta \;\bar R_\mu (x) + \bar \theta \;R_\mu (x)
+ i \;\theta \;\bar\theta \;S_\mu (x), \nonumber\\
{\cal F} (x, \theta, \bar \theta) &=& C (x) + i \;\theta \; \bar B_1 (x) + i \;\bar\theta \;B_1 (x)
+ i \;\theta \;\bar\theta\; s(x), \nonumber\\
\bar {\cal F} (x, \theta, \bar \theta) &=& \bar C (x) + i \;\theta \;\bar B_2 (x) + i\; \bar\theta \;B_2 (x)
+ i \;\theta \;\bar\theta \;\bar s (x), 
\end{eqnarray}
where the secondary fields ($\bar R_\mu (x), R_\mu (x), s(x), \bar s(x)$) are fermionic and the
other secondary fields ($S_\mu (x), B_1 (x), \bar B_1 (x), B_2 (x), \bar B_2 (x)$) are bosonic in nature.
These secondary fields are determined in terms of the basic and auxiliary fields of the 4D non-Abelian
1-form gauge theory by exploiting the mathematical power of the HC.

Under the celebrated HC, the $SU(N)$ gauge-invariant kinetic term
($ - \frac{1}{4}\; F^{\mu\nu} \cdot F_{\mu\nu}$) of the 4D non-Abelian gauge theory is required to
remain invariant when we generalize the 4D local non-Abelian theory onto the (4, 2)-dimensional
supermanifold in terms of the superfields. In other words, 
the super 2-form $\tilde {\cal F}^{(2)} = \tilde d \tilde A^{(1)} 
+ i \; \tilde A^{(1)} \wedge \tilde A^{(1)} \equiv \frac{1}{2!} (dZ^M \wedge dZ^N)\; \tilde F_{MN}$, 
defined on the (4, 2)-dimensional supermanifold with
the following inputs
\begin{eqnarray}
\tilde d = d Z^M \partial_M &\equiv& dx^\mu \;\partial_\mu + d \theta \;\partial_\theta
+ d \bar \theta \;\partial_{\bar\theta}, \qquad  
\partial_M = (\partial_\mu, \partial_\theta, \partial_{\bar\theta})\nonumber\\
\tilde A^{(1)} = d Z^M A_M &\equiv& dx^\mu \;{\cal B}_\mu (x,\theta,\bar\theta) + d \theta \; 
\bar {\cal F} (x,\theta,\bar\theta)
+ d \bar \theta\; {\cal F} (x,\theta,\bar\theta), 
\end{eqnarray} 
is equated to the ordinary 2-form $F^{(2)} = d A^{(1)} + i\; A^{(1)} \wedge A^{(1)}$
in the HC.
The latter defines the ordinary curvature tensor $F_{\mu\nu} = \partial_\mu A_\nu
- \partial_\nu A_\mu + i \; [A_\mu, A_\nu]$.
In the above, the super multiplet  $A_M = ({\cal B}_\mu, {\cal F}, \bar {\cal F})$ is defined on the
(4, 2)-dimensional supermanifold which is characterized in terms of 
the superspace coordinates $Z^M = (x^\mu, \theta, \bar\theta)$.


In the HC, all the Grassmannian components of the super curvature
$\tilde {\cal F}_{MN}$ are set equal to zero. This requirement leads to the following relationships [1,2]
\begin{eqnarray}
&& R_\mu = D_\mu C, \quad \bar R_\mu = D_\mu \bar C, \quad B_1 = - \frac{i}{2} (C \times C),
\quad  s = - (\bar B_1 \times C), \nonumber\\
&& S_\mu = D_\mu B_2 + i (D_\mu C \times \bar C) \equiv - D_\mu \bar B_1 - i (C \times D_\mu \bar C), \nonumber\\
&& \bar B_2 = - \frac{i}{2} (\bar C \times \bar C), \quad
\bar B_1 + B_2 = - i (C \times \bar C), \quad \bar s = - (B_2 \times \bar C).
\end{eqnarray}
If we make the identifications: $ \bar B_1 = \bar B, B_2 = B$, the above Curci-Ferrari restriction 
$\bar B_1 + B_2 = - i (C \times \bar C)$  changes to its well-known form $ B + \bar B = - i (C \times \bar C)$. 
Plugging in the above relationships in the expansions (4), we obtain the following expressions for the superfields
along the Grassmannian directions of the supermanifold, namely;
\begin{eqnarray}
{\cal B}^{(h)}_\mu (x, \theta, \bar \theta) &=& A_\mu (x) + \theta \; (D_\mu \bar C(x)) + \bar \theta \;
(D_\mu C(x))
+ \theta \;\bar\theta \;[i\;D_\mu B (x) - (D_\mu C \times C)(x)], \nonumber\\
{\cal F}^{(h)} (x, \theta, \bar \theta) &=& C (x) + \theta \; (i \bar B (x)) + \bar\theta \; 
\Bigl [\frac{1}{2} (C \times C)(x) \Bigr ]
+ \theta \;\bar\theta\;[-i\; (\bar B \times C) (x)], \nonumber\\
\bar {\cal F}^{(h)} (x, \theta, \bar \theta) &=& \bar C (x) + \theta \;\Bigl [\frac{1}{2} (\bar C \times \bar C) (x)
\Bigr ] +  \bar\theta \; (i B (x))
+ \theta \;\bar\theta \;[ - i\; (B \times \bar C)(x)], 
\end{eqnarray}
which can be expressed in terms the 
off-shell nilpotent ($s_{(a)b}^2 = 0$) (anti-) BRST symmetry transformations 
$ s_{(a)b}$ for the non-Abelian 1-form gauge theory
as follows\footnote{The full off-shell nilpotent 
transformations $s_{(a)b}$ (cf. (16) below) are absolutely anticommuting on
a surface described by the Curci-Ferrari field equation [$B + \bar B + i (C \times \bar C) = 0$] in the 4D spacetime manifold.}
\begin{eqnarray}
{\cal B}^{(h)}_\mu (x, \theta, \bar \theta) &=& A_\mu (x) + \theta \; (s_{ab} A_\mu (x)) + \bar \theta \;
(s_b A_\mu (x))
+ \theta \;\bar\theta \;(s_b s_{ab} A_\mu (x)), \nonumber\\
{\cal F}^{(h)} (x, \theta, \bar \theta) &=& C (x) + \theta \; (s_{ab} C(x)) + \bar\theta \; 
(s_b C(x))
+ \theta \;\bar\theta\;(s_b s_{ab} C(x)), \nonumber\\
\bar {\cal F}^{(h)} (x, \theta, \bar \theta) &=& \bar C (x) + \theta \;(s_{ab} \bar C(x)) +  \bar\theta \; 
(s_b \bar C (x))
+ \theta \;\bar\theta \;(s_b s_{ab} \bar C(x)), 
\end{eqnarray}
where the superscript $(h)$ on the superfields denotes the expansions 
of the superfields after the application of the horizontality condition.

The spacetime component
of the super curvature tensor $\tilde {\cal F}_{MN}$ is $\tilde {\cal F}^{(h)}_{\mu\nu} (x,\theta,\bar\theta)
= \partial_\mu {\cal B}^{(h)}_\nu - \partial_\nu {\cal B}^{(h)}_\mu 
+ i [ {\cal B}^{(h)}_\mu, {\cal B}^{(h)}_\nu ]$. This can be written, using the expansion
for ${\cal B}^{(h)}_\mu (x,\theta,\bar\theta)$ in (7), as
\begin{eqnarray}
\tilde {\cal F}^{(h)}_{\mu\nu} (x,\theta,\bar\theta) &=& F_{\mu\nu} - \theta \;(F_{\mu\nu} \times \bar C)
- \bar \theta \;(F_{\mu\nu} \times  C) + \theta \bar \theta \;[(F_{\mu\nu} \times C) \times \bar C - i\; 
F_{\mu\nu} \times B].
\end{eqnarray}
The above expression does imply clearly that the kinetic term remains invariant under the horizontality
condition (i.e. $- \frac{1}{4} \tilde F^{\mu\nu (h)} (x,\theta,\bar\theta) 
\cdot \tilde F^{(h)}_{\mu\nu} (x,\theta,\bar\theta)
= - \frac{1}{4} F^{\mu\nu} \cdot F_{\mu\nu}$). 

\section{Off-shell nilpotent transformations for non-Abelian 2-form gauge and 1-form auxiliary fields}

Exploiting  (3), it can be checked that
$\delta_g \bigl (B_{\mu\nu} \cdot F_{\eta\kappa} \bigr ) = 0, 
\delta_g \bigl (K_{\mu} \cdot F_{\nu\eta} \bigr ) = 0.$
Thus, we propose the following gauge-invariant restrictions (GIRs) in terms of the (super)fields
\begin{eqnarray}
 \tilde {\cal B}_{\mu\nu} (x,\theta,\bar\theta) \cdot \tilde {\cal F}^{(h)}_{\eta\kappa} (x,\theta,\bar\theta)
&=& B_{\mu\nu} (x) \cdot F_{\eta\kappa} (x), \nonumber\\
\tilde {\cal K}_{\mu} (x,\theta,\bar\theta) \cdot \tilde {\cal F}^{(h)}_{\nu\eta} (x,\theta,\bar\theta)
&=& K_{\mu} (x) \cdot F_{\nu\eta} (x),
\end{eqnarray}
as analogues of the horizontality condition ($\tilde {\cal F}^{(2)} = F^{(2)}$).
The expansions of the superfields $\tilde {\cal B}_{\mu\nu} (x,\theta,\bar\theta)$ and
$\tilde {\cal K}_{\mu} (x,\theta,\bar\theta)$ on the (4, 2)-dimensional supermanifold are
\begin{eqnarray}
\tilde {\cal B}_{\mu\nu} (x, \theta, \bar \theta) &=& B_{\mu\nu} (x) + \theta \; \bar R_{\mu\nu} (x)) + \bar \theta \;
R_{\mu\nu} (x)
+ i\;\theta \;\bar\theta \; S_{\mu\nu} (x), \nonumber\\
\tilde {\cal K}_\mu (x, \theta, \bar \theta) &=& K_\mu (x) + \theta \; \bar P_\mu (x) + \bar\theta \; 
P_\mu (x) + i\;\theta \;\bar\theta\;Q_\mu (x), 
\end{eqnarray}
where the secondary fields ($R_{\mu\nu}, \bar R_{\mu\nu}, P_\mu, \bar P_\mu$) are fermionic 
and ($S_{\mu\nu}, Q_\mu$) are bosonic in nature. These secondary fields would be determined
by exploiting the above restrictions (10) where the HC plays
a decisive role, too, in a subtle manner.

It is straightforward to check that the following relationships ensue from (10):
\begin{eqnarray}
&& R_{\mu\nu} = - (B_{\mu\nu} \times C) \;  \bar R_{\mu\nu} = - (B_{\mu\nu} \times \bar C), \;
S_{\mu\nu} = - (B_{\mu\nu} \times B) - i [(B_{\mu\nu} \times C) \times \bar C], \nonumber\\
&& P_\mu  = - (K_\mu \times C), \quad \bar P_\mu = - (K_\mu \times \bar C), \quad
Q_\mu = - (K_{\mu} \times B) - i \;[(K_{\mu} \times C) \times \bar C].
\end{eqnarray}
The expansions, that emerge after the application of the gauge-invariant restrictions, are
\begin{eqnarray}
\tilde {\cal B}^{(g)}_{\mu\nu} (x, \theta, \bar \theta) &=& B_{\mu\nu} (x) - \theta \; 
[(B_{\mu\nu} \times \bar C) (x)] - \bar \theta \; [(B_{\mu\nu} \times C) (x)] \nonumber\\
&+& \theta \;\bar\theta \; [ \{(B_{\mu\nu} \times C) \times \bar C - i\; B_{\mu\nu} \times B \} (x)], \nonumber\\
&\equiv & B_{\mu\nu} (x) + \theta\; (s_{ab} B_{\mu\nu} (x)) + \bar\theta\; (s_b B_{\mu\nu} (x))
+ \theta\;\bar\theta\; (s_b s_{ab} B_{\mu\nu} (x)), \nonumber\\
\tilde {\cal K}^{(g)}_\mu (x, \theta, \bar \theta) &=& K_\mu (x) - \theta \; [(K_\mu \times \bar C) (x)]
 - \bar\theta \;  [(K_\mu \times  C) (x)] \nonumber\\
&+& \theta \;\bar\theta\; [ \{(K_\mu \times C) \times \bar C - i \;K_\mu \times B \} (x)], \nonumber\\
&\equiv & K_{\mu} (x) + \theta\; (s_{ab} K_{\mu} (x)) + \bar\theta\; (s_b K_{\mu} (x))
+ \theta\;\bar\theta\; (s_b s_{ab} K_{\mu} (x)), 
\end{eqnarray}
where the superscript $(g)$ denotes the super expansions obtained after the application of GIRs.
From the preceding discussions, it is clear that we have obtained all the off-shell nilpotent
(anti-) BRST transformations for the basic fields ($B_{\mu\nu}, A_\mu$), auxiliary field ($K_\mu$)
and the (anti-) ghost fields $(\bar C) C$ by exploiting the geometrical superfield formalism. 
As pointed out earlier, the (anti-) BRST symmetry transformations for the auxiliary fields follow
from the requirements of the properties of nilpotency and anticommutativity.

From the gauge-invariant restrictions (10) and super expansions in (9) and (13), it is
clear that the topological term in (1) remains invariant when we generalize the 4D theory
onto the (4, 2)-dimensional supermanifold. As a consequence, we have the
equality
\begin{eqnarray}
\frac{m}{4}\; \varepsilon^{\mu\nu\eta\kappa} \; \tilde {\cal B}^{(g)}_{\mu\nu} \;(x,\theta,\bar\theta)
\; \cdot \; \tilde {\cal F}^{(h)}_{\eta\kappa} \;(x,\theta,\bar\theta)
= \frac{m}{4}\; \varepsilon^{\mu\nu\eta\kappa}\; B_{\mu\nu} (x) \cdot F_{\eta\kappa} (x).
\end{eqnarray}
It is worth pointing out that the above
equality shows that, ultimately, the l.h.s. of (14) is independent of 
the Grassmannian varibales. In an exactly similar fashion,
it can be checked that the following expression for the super 
curvature tensor\footnote{The superscripts ${(g, h)}$, on the 
compensated super curvature tensor $\tilde {\cal H}^{(g,h)}_{\mu\nu\eta} \;(x,\theta,\bar\theta)$, 
denote the incorporation of the constituent superfields
(i.e. $\tilde {\cal B}^{(g)}_{\mu\nu}, \tilde {\cal K}^{(g)}_\mu, \tilde {\cal F}^{(h)}_{\mu\nu}$),
that have been obtained after the application of the HC and GIRs. The latter are found to be
complementary and consistent with each-other.} 
($\tilde {\cal H}^{(g,h)}_{\mu\nu\eta} \;(x,\theta,\bar\theta)$)
\begin{eqnarray}
&&\tilde {\cal H}^{(g,h)}_{\mu\nu\eta}\; (x, \theta, \bar \theta) = H_{\mu\nu\eta} (x) - \;\theta \; 
[(H_{\mu\nu\eta} \times \bar C) (x)] - \;\bar \theta \; [(H_{\mu\nu\eta} \times  C) (x)] \nonumber\\
&& + \;  \theta \;\bar\theta \; [ \{ (H_{\mu\nu\eta} \times C) \times \bar C - i \; H_{\mu\nu\eta} \times B \} (x)]
\nonumber\\
&& \equiv  H_{\mu\nu\eta} (x) + \theta\; (s_{ab} H_{\mu\nu\eta} (x)) + \bar\theta\; (s_b H_{\mu\nu\eta} (x))
+ \theta\;\bar\theta\; (s_b s_{ab} H_{\mu\nu\eta} (x)),
\end{eqnarray}
implies that
$\frac{1}{12} \tilde {\cal H}^{\mu\nu\eta(g,h)} (x,\theta,\bar\theta) \cdot
\tilde {\cal H}^{(g,h)}_{\mu\nu\eta} (x,\theta,\bar\theta) = 
\frac{1}{12} H^{\mu\nu\eta} (x) \cdot H_{\mu\nu\eta} (x).$
In other words, the l.h.s. of the above expression (that is defined on terms of
the superfields located on the (4, 2)-dimensional supermanifold) is 
independent\footnote{It
is interesting to note that the l.h.s. of the above equality remains independent
of the Grassmannian variables $\theta$ and $\bar\theta$ when we exploit the full expansion (15).
In fact, the coefficients of $\theta$, $\bar\theta$ and $\theta\bar\theta$ of the l.h.s.
turn out to be zero by use of the totally antisymmetric properties of the structure constants
$f^{abc}$.}  
of the Grassmannian variables of the superspace coordinates 
$Z^M = (x^\mu,\theta, \bar\theta)$. 

\section{Coupled Lagrangian densities and their invariance}

It can be checked from the action (corresponding to the 
starting Lagrangian density (1)) and the following off-shell nilpotent
(anti-) BRST symmetry transformations
\begin{eqnarray}
&& s_b A_\mu = D_\mu C, \qquad s_b C = \frac{1}{2} (C \times C), \qquad
s_b \bar C = i B,  \qquad  s_b B = 0, \nonumber\\
&& s_b \bar B = - (\bar B \times C), \quad s_b F_{\mu\nu} = - (F_{\mu\nu} \times C),
\quad s_b H_{\mu\nu\eta} = - (H_{\mu\nu\eta} \times C), \nonumber\\
&& s_b B_{\mu\nu} = - (B_{\mu\nu} \times C), \qquad s_b K_\mu = - (K_\mu \times C), \nonumber\\
&& s_{ab} A_\mu = D_\mu \bar C, \quad s_{ab} \bar C = \frac{1}{2} (\bar C \times \bar C), \quad
s_{ab} C = i \bar B,  \quad  s_{ab} \bar B = 0, \nonumber\\
&& s_{ab} B = - (B \times \bar C), \quad s_{ab} F_{\mu\nu} = - (F_{\mu\nu} \times \bar C),
\quad s_{ab} H_{\mu\nu\eta} = - (H_{\mu\nu\eta} \times \bar C), \nonumber\\
&& s_{ab} B_{\mu\nu} = - (B_{\mu\nu} \times \bar C), \qquad s_{ab} K_\mu = - (K_\mu \times \bar C),
\end{eqnarray}
that the mass dimensions of the fields of the theory,
in natural units $\hbar = c = 1$, are: $ [ A_\mu ] = [ B_{\mu\nu} ] = [C] = [\bar C] = [M],\;
[K_\mu] = [0],\; [F_{\mu\nu}] = [H_{\mu\nu\eta}] = [B] = [\bar B] = [M]^2$.

As a consequence of the above observations, the expressions
for the (anti-) BRST invariant coupled Lagrangian densities 
can be written as follows
\begin{eqnarray}
{\cal L}_B &=& - \frac{1}{4}\; F^{\mu\nu} \cdot F_{\mu\nu} + \frac{1}{12} \;H^{\mu\nu\eta} \cdot H_{\mu\nu\eta}
+ \frac{m}{4} \;\varepsilon_{\mu\nu\eta\kappa} B^{\mu\nu} \cdot F^{\eta\kappa} \nonumber\\
&+& s_b s_{ab} \;\Bigl (\frac{1}{4} B^{\mu\nu} \cdot B_{\mu\nu} + \frac{i}{2} \; A^\mu \cdot A_\mu
+ C \cdot \bar C \Bigr ), \nonumber\\
{\cal L}_{\bar B} &=& - \frac{1}{4}\; F^{\mu\nu} \cdot F_{\mu\nu} + \frac{1}{12} \;H^{\mu\nu\eta} \cdot H_{\mu\nu\eta}
+ \frac{m}{4} \;\varepsilon_{\mu\nu\eta\kappa} B^{\mu\nu} \cdot F^{\eta\kappa} \nonumber\\
&-& s_{ab} s_b \;\Bigl (\frac{1}{4} B^{\mu\nu} \cdot B_{\mu\nu} + \frac{i}{2} \; A^\mu \cdot A_\mu
+ C \cdot \bar C \Bigr ).
\end{eqnarray}
It should be noted that, in the above parenthesis, we have chosen the combinations of fields
that have, in totality, mass dimension equal to {\it two} and ghost number equal to {\it zero}. As a result,
we have the following coupled Lagrangian densities 
\begin{eqnarray}
{\cal L}_B &=& - \frac{1}{4}\; F^{\mu\nu} \cdot F_{\mu\nu} + \frac{1}{12} \;H^{\mu\nu\eta} \cdot H_{\mu\nu\eta}
+ \frac{m}{4} \;\varepsilon_{\mu\nu\eta\kappa} B^{\mu\nu} \cdot F^{\eta\kappa} \nonumber\\
&+& B \cdot (\partial_\mu A^\mu) + \frac{1}{2}\; \Bigl (B \cdot B + \bar B \cdot \bar B \Bigr )
- i \;\partial_\mu \bar C \cdot D^\mu C, \nonumber\\
{\cal L}_{\bar B} &=& - \frac{1}{4}\; F^{\mu\nu} \cdot F_{\mu\nu} + \frac{1}{12} \;H^{\mu\nu\eta} \cdot H_{\mu\nu\eta}
+ \frac{m}{4} \;\varepsilon_{\mu\nu\eta\kappa} B^{\mu\nu} \cdot F^{\eta\kappa} \nonumber\\
&-& \bar B \cdot (\partial_\mu A^\mu) 
+ \frac{1}{2}\; \Bigl (B \cdot B + \bar B \cdot \bar B \Bigr )
- i \;D_\mu \bar C \cdot \partial^\mu C.
\end{eqnarray} 
It can be checked that the Lagrangian densities ${\cal L}_B$ and ${\cal L}_{\bar B}$
transform under the off-shell nilpotent BRST and anti-BRST symmetry transformations (cf. (16)) as  
\begin{eqnarray}
&& s_b {\cal L}_B = \partial_\mu \bigl [ B \cdot D^\mu C \bigr ], \quad
s_{ab} {\cal L}_{\bar B} = - \partial_\mu \bigl [ \bar B \cdot D^\mu \bar C \bigr ] \quad
s_b {\cal L}_{\bar B} = - \partial_\mu \bigl [\bar B \cdot \partial^\mu C \bigr ] 
+ D_\mu \bigl [ B + \bar B \nonumber\\ && + i (C \times \bar C) \bigr ] \cdot \partial^\mu C, \quad
s_{ab} {\cal L}_B = \partial_\mu \bigl [B \cdot \partial^\mu \bar C \bigr ] - 
D_\mu \bigl [ B + \bar B + i (C \times \bar C) \bigr ] \cdot \partial^\mu \bar C.
\end{eqnarray}
Thus, the action corresponding to the above Lagrangian densities remains
invariant.

The 4D coupled
Lagrangian densities (17) can be generalized onto (4, 2)-dimensional supermanifold and can
be expressed in terms of the superfields obtained after the applications of HC and GIRs. These 
super Lagrangian densities, in full blaze of glory, are
\begin{eqnarray}
\tilde {\cal L}_B &=& - \frac{1}{4}\; \tilde {\cal F}^{\mu\nu (h)} \cdot \tilde {\cal F}^{(h)}_{\mu\nu} 
+ \frac{1}{12} \;\tilde {\cal H}^{\mu\nu\eta (g,h)} \cdot \tilde {\cal H}^{(g,h)}_{\mu\nu\eta}
+ \frac{m}{4} \;\varepsilon_{\mu\nu\eta\kappa} \tilde {\cal B}^{\mu\nu (g)} \cdot \tilde {\cal F}^{\eta\kappa (h)} \nonumber\\
&+& \frac{\partial}{\partial \bar\theta}\; \frac{\partial}{\partial \theta}\; 
\Bigl (\frac{1}{4} \tilde {\cal B}^{\mu\nu (g)} \cdot \tilde {\cal B}^{(g)}_{\mu\nu} 
+ \frac{i}{2} \; {\cal B}^{\mu (h)} \cdot  {\cal B}^{(h)}_\mu
+  \tilde {\cal F}^{(h)} \cdot \bar {\cal F}^{(h)}  \Bigr ), \nonumber\\
\tilde {\cal L}_{\bar B} &=& - \frac{1}{4}\; \tilde {\cal F}^{\mu\nu (h)} \cdot \tilde {\cal F}^{(h)}_{\mu\nu} 
+ \frac{1}{12} \;\tilde {\cal H}^{\mu\nu\eta (g,h)} \cdot \tilde {\cal H}^{(g,h)}_{\mu\nu\eta}
+ \frac{m}{4} \;\varepsilon_{\mu\nu\eta\kappa} \tilde {\cal B}^{\mu\nu (g)} \cdot 
\tilde {\cal F}^{\eta\kappa (h)} \nonumber\\
&-& \frac{\partial}{\partial \theta}\; \frac{\partial}{\partial \bar \theta}\; 
\Bigl (\frac{1}{4} \tilde {\cal B}^{\mu\nu (g)} \cdot \tilde {\cal B}^{(g)}_{\mu\nu} 
+ \frac{i}{2} \; {\cal B}^{\mu (h)} \cdot  {\cal B}^{(h)}_\mu
+  \tilde {\cal F}^{(h)} \cdot  {\bar {\cal F}}^{(h)}  \Bigr ).
\end{eqnarray}
The BRST and anti-BRST invariance of equation (19) can be translated into the language of the super
Lagrangian densities (20) and the operation on them by the 
Grassmannian partial derivatives as:
$(\partial/ \partial \bar\theta) \tilde {\cal L}_B = 0, \; (\partial /\partial \bar\theta) 
\tilde {\cal L}_{\bar B} = 0, \;(\partial /\partial \theta) \tilde {\cal L}_B = 0,\;
(\partial /\partial \theta) \tilde {\cal L}_{\bar B} = 0.$
Thus, within the framework of the geometrical superfield formulation, we have captured the (anti-) BRST invariance
of the 4D coupled Lagrangian densities in a simple manner. 

\section{ Conclusions}

One of the key observations of our present investigation is to obtain the compelling theoretical reasons
to go beyond the application of the HC in the context of superfield formulation of purely
free $p$-form ($p = 1, 2, 3....$) gauge theories (where there is no interaction with matter fields). 
As it turns out, the GIRs on the superfields complement the application of the HC in the sense that
we derive {\it all} the off-shell nilpotent (anti-) BRST symmetry transformations for the present
4D topologically massive non-Abelian gauge theory. We have exploited the GIRs in the context
of (non-) Abelain 1-form gauge theory as well [2,3]. The distinct difference, however, is
that, in all such theories [2,3], there is presence of matter fields. 
It is worth pointing out that we have tapped only the {\it usual} ``scalar''
gauge symmetry transformations for
our BRST analysis and have ignored the ``vector'' gauge symmetry transformations
(cf. Sec. 1, for some concise remarks on it). It would be very interesting to exploit both these gauge symmetries
{\it together} for the BRST analysis within the framework of superfield approach to our 4D
topologically massive non-Abelian model. \\

\noindent
{\bf Acknowledgements:}
Financial support from DST, Government of India,  
under the SERC project sanction grant No: - SR/S2/HEP-23/2006, is gratefully acknowledged.

\end{document}